\documentclass[useAMS,usenatbib]{mn2e}
\usepackage[dvips]{graphics}

\newcommand{\bea}{\begin{eqnarray}}
\newcommand{\eea}{\end{eqnarray}}
\newcommand{\beq}{\begin{equation}}
\newcommand{\eeq}{\end{equation}}
\newcommand{\ve}[1]{{\mathbf #1}}
\newcommand{\bP}{\mathbf{P}}
\newcommand{\bF}{\mathbf{F}}
\newcommand{\bZ}{\mathbf{Z}}
\newcommand{\bI}{\mathbf{I}}

\newcommand{\bC}{\mathbf{C}}
\newcommand{\fknee}{f_\mathrm{kn}}
\newcommand{\fspin}{f_\mathrm{sp}}
\newcommand{\fmin}{f_\mathrm{min}}
\newcommand{\fmax}{f_\mathrm{max}}
\newcommand{\fsample}{f_\mathrm{s}}
\newcommand{\muK}{$\mu$K}
\newcommand{\Planck}{{\it Planck }}
\newcommand{\Madam}{{\small MADAM}}

\title{Madam - a map-making method for CMB experiments}

\author{E. Keih\"anen$^{1,2}$\thanks{E-mail: elina.keihanen@helsinki.fi}, 
H. Kurki-Suonio$^1$, and T. Poutanen$^{2}$ \\
$^1$University of Helsinki, Department of Physical Sciences,
P.O. Box 64, FIN-00014, Helsinki, Finland \\
$^2$Helsinki Institute of Physics, P.O. Box 64, FIN-00014, Helsinki, Finland \\
}

\begin{document}

\maketitle

\begin{abstract}
We present a new map-making method for CMB measurements. 
The method is based on the destriping technique, but it also 
utilizes information about the noise spectrum. The low-frequency 
component of the instrument noise stream is modelled as a superposition 
of a set of simple base functions, whose amplitudes are determined 
by means of maximum-likelihood analysis, involving the covariance matrix 
of the amplitudes. 
We present simulation results with $1/f$ noise and show a reduction 
in the residual noise with respect to ordinary destriping.  
This study is related to \Planck LFI activities.
\end{abstract}



\section{Introduction}

Construction of the CMB map from time-ordered data (TOD) is 
an important part of the data analysis of CMB experiments.
Future space missions like \Planck
\footnote{http://astro-estec.esa.nl/SA-general/Projects/Planck}
present new challenges for the data analysis. 
The amount of data \Planck produces is far larger than that 
of any earlier experiments.

The destriping technique
(Burigana et al. 1997, Delabrouille 1998, Maino et al. 1999, 2002, 
Keih\"anen et al. 2004)
provides an efficient map-making method for large data sets.
The method is non-optimal in accuracy but fast and stable.
Other methods have been developed which aim at finding 
the optimal minimum-variance map for \Planck-like data (e.g.
Natoli et al. 2001, Dor\'e et al. 2001, Borrill 1999). 

In this paper we present a new map-making method for CMB 
measurements, called
\Madam\ ({\bf Ma}p-making through {\bf D}estriping for 
{\bf A}nisotropy {\bf M}easurements).
The method is built on the destriping technique 
but unlike ordinary destriping, it also utilizes information about 
the noise spectrum. The aim is to improve the accuracy of the output 
map as compared to destriping, while still keeping, at least partly, 
the speed and stability of the destriping method.

The basic idea of the method is the following.
The low-frequency component of the instrument noise in the TOD 
is modelled as a superposition of simple base 
functions, whose amplitudes are determined by means of 
maximum-likelihood analysis, involving the covariance matrix of the 
amplitudes. The covariance matrix is computed from the noise spectrum, 
assumed to be known. 
 
This paper is organized as follows. In Section 2 we go through the 
maximum-likelihood analysis that forms the basis of our map-making method.
We describe the map-making algorithm in Section 3. In Section 4
we consider the covariance matrix of component amplitudes. 
Some technical calculations related to this section are presented 
in Appenxix A.
We give results from numerical simulations in Section 5 and 
present our conclusions in Section 6.


\section{Map-making problem}

\subsection{Maximum-likelihood analysis of the destriping problem}
\label{sec:ml}

In the following we present a maximum-likelihood analysis on which
our map-making method is based. The analysis is similar to that 
presented in \cite{Keihanen04}, the main difference being, that here 
we include the covariance of the correlated noise component.

We write the time-ordered data (TOD) stream as
 \beq
    \ve{y} = \bP\ve{m} +\ve{n}'. \label{tod1}
\eeq
Here the first term presents the CMB signal and the second 
term presents noise.
Vector $\ve{m}$ presents the pixelized CMB map and pointing
matrix $\bP$ spreads it into TOD. 

We divide the noise contribution into a correlated noise component 
and white noise, and model the correlated part as a linear combination
of some orthogonal base functions,
 \beq
    \ve{n}' = \bF\ve{a} +\ve{n}. \label{noisetod}
\eeq
Vector $\ve a$ contains the unknown amplitudes of the base functions
and matrix $\bF$ spreads them into TOD. Each column of matrix $\bF$
contains the values of the corresponding base function along the TOD. 
Assuming that the white noise
component and the correlated noise component are independent,
the total noise covariance is given by
\beq
   \bC_t = \langle\ve n'(\ve n')^T\rangle
       = \bF\bC_a\bF^T +\bC_n   \label{totcov}
\eeq
where $\bC_n=\langle\ve n\ve n^T\rangle$ is the white noise
covariance, $\bC_a=\langle\ve a\ve a^T\rangle$ is the covariance 
matrix for the component amplitudes $\ve{a}$, and $\langle x\rangle$
denotes the expectation value of quantity $x$.

Our aim is to find the maximum-likelihood estimate of 
$\ve m$ and $\ve a$ simultaneously, for given data $\ve y$.
We consider the likelihood of the data, which by the chain rule 
of probability theory can be written as
\beq
     P(\ve y) = P(\ve y|\ve m,\ve a)P(\ve a|\ve m)P(\ve m). 
     \label{chain}
\eeq
Here $P(a|b)$ denotes the conditional probability 
of $a$ under condition $b$.
Now $P(\ve m)$ is constant, since we are treating the underlying 
CMB sky as deterministic (we associate no probability distribution 
to it).
The probability distribution of $\ve a$ is independent of the map 
so that 
$P(\ve a|\ve m)=P(\ve a)$. We assume gaussianity and write
\beq
    P(\ve a) = (2\pi\det\bC_a)^{-1/2}
               \exp(-\frac12\ve a^T\bC_a^{-1}\ve a).
\eeq
With $\ve m$ and $\ve a$ fixed, the likelihood of the data is
given by the white noise distribution
\beq
    P(\ve y|\ve m,\ve a) = 
    (2\pi\det\bC_n)^{-1/2}
    \exp(-\frac 12 \ve n^T
                   \bC_n^{-1}\ve n)
\eeq
where now $\ve n=\ve y-\bF\ve a-\bP\ve m$.
The white noise covariance $\ve C_n$ 
is assumed to be diagonal, but not necessarily uniform.

Maximizing the likelihood (\ref{chain}) is equivalent to mimimizing
the inverse of its logarithm. We obtain the chi-square 
minimization function
\bea
      \chi^2 &=& -2\ln P(\ve y) 
        = -2\ln (P(\ve y|\ve m,\ve a)P(\ve a))\hfill  \nonumber \\
             &=&(\ve{y}-\bF\ve{a}-\bP\ve{m})^T
                \bC_n^{-1}(\ve{y}-\bF\ve{a}-\bP\ve{m})\nonumber\\
             && +\ve{a}^T\bC_a^{-1}\ve{a} + \hbox{constant}
      \label{chi1}
\eea

We want to minimize $(\ref{chi1})$ with respect to both $\ve{a}$ and
 $\ve{m}$.
Minimization with respect to $\ve{m}$ gives 
\beq   
          \ve{m}= (\bP^T\bC_n^{-1}\bP)^{-1}
           \bP^T\bC_n^{-1}(\ve{y}-\bF\ve{a}) \, .  \label{mlmap}
\eeq
Substituting Eq.~(\ref{mlmap}) back into Eq.~(\ref{chi1}) we get
the chi-square into the form
\beq
    \chi^2 = (\ve{y}-\bF\ve{a})^T \bZ^T\bC_n^{-1}\bZ(\ve{y}-\bF\ve{a})
    +\ve{a}^T\bC_a^{-1}\ve{a}, \label{chin}
\eeq
where
\beq
      \bZ = \bI-\bP(\bP^T\bC_n^{-1}\bP)^{-1} \bP^T\bC_n^{-1} \, .
\eeq
Here $\bI$ denotes the unit matrix.

We minimize $\chi^2$ with respect to $\ve{a}$, to obtain 
an estimate for the amplitude vector $\ve a$. 
The solution is given by
\beq
    (\bF^T\bC_n^{-1}\bZ\bF+\bC_a^{-1})\ve{a} 
     = \bF^T\bC_n^{-1}\bZ\ve{y}.  \label{linc}
\eeq
Here we have used the property $\bZ^T\bC_n^{-1}\bZ=\bC_n^{-1}\bZ$.

The \Madam\ algorithm uses the conjugate gradient technique 
to solve vector $\ve{a}$ from (\ref{linc}). 
Note that the matrix on the left-hand side 
is symmetric.
An estimate for the CMB map can then be computed using 
Eq. (\ref{mlmap}).

\subsection{Comparison to the minimum-variance solution}
\label{sec:gls}

If the underlying CMB map is treated as deterministic, noise is 
Gaussian, and its statistical properties are known, the optimal 
minimum-variance map is given by
\beq
      \ve{m}= (\bP^T\bC_t^{-1}\bP)^{-1} \bP^T \bC_t^{-1}\ve{y}
      \label{glsmap}
\eeq
where $\bC_t$ is the covariance of the noise TOD.

In the following we show that if the total noise covariance is of the 
form (\ref{totcov}), the map estimate given by Eqs. (\ref{mlmap})
and (\ref{linc}) equals the minimum-variance map (\ref{glsmap}).

We develop the solution (\ref{linc}) into Taylor 
series as
\beq
    \ve a = (\bC_a -\bC_a\bF^T\bC_n^{-1}\bZ\bF\bC_a+\ldots)
       \bF^T\bC_n^{-1}\bZ\ve{y} .
\eeq
We write $\ve y-\bF\ve a$ out and recollapse the resulting expansion, 
to get
\beq
   \ve y-\bF\ve a =
       (\bI+\bF\bC_a\bF^T\bC_n^{-1}\bZ)^{-1}\ve{y}.   \label{yfa} 
\eeq
We now use Eq. (\ref{totcov}) and write
 $\bF\bC_a\bF^T\bC_n^{-1}=\bC_t\bC_n^{-1}-\bI$. 
Using this and writing $\bZ$ out we arrive at
\bea   
   \lefteqn{ \bP^T\bC_n^{-1}(\ve y-\bF\ve a) }   \\
     &=& \bP^T[\bC_t -(\bC_t\bC_n^{-1}-\bI)
          \bP(\bP^T\bC_n^{-1}\bP)^{-1}\bP^T ]^{-1}\ve{y} \nonumber \\
     &=& [\bI -\bP^T(\bC_n^{-1}-\bC_t^{-1})
          \bP(\bP^T\bC_n^{-1}\bP)^{-1}]^{-1}\bP^T\bC_t^{-1}\ve{y}.
          \nonumber 
\eea
In the last equality we have taken $\bC_t$ out from the left
and used the identity 
$\mathbf{A}(\bI+\mathbf{BA})^{-1}=(\bI+\mathbf{AB})^{-1}\mathbf{A}$,
which is easily verified by expanding both sides as Taylor series.
The \Madam\ solution for the map (Eq. (\ref{mlmap})) then becomes
\beq
     \ve{m} 
    = [\bP^T\bC_n^{-1}\bP-\bP^T(\bC_n^{-1}-\bC_t^{-1})\bP]^{-1}
        \bP^T\bC_t^{-1}\ve{y}    
\eeq
which readily simplifies into (\ref{glsmap}).

If the chosen base function set accurately models 
the correlated noise component, 
the CMB map estimate given by \Madam\ equals 
the minimum-variance solution.
This is necessarily true at the limit where the number of 
base functions $L$ per ring approaches the number of samples $n$,
since the base functions then form a complete orthogonal basis.
In practice, however, it is not possible to use that many base 
functions, since both the required memory and CPU time increase
with increasing $L$ so that the method finally becomes unfeasible.


\section{Implementation}

\subsection{Map-making algorithm}
\label{sec:algorithm}

Equations (\ref{mlmap}) and (\ref{linc}) form the basis of the 
\Madam\ map-making method.
In this Section we consider the implementation of 
the method.

Our starting point is a \Planck-like scanning strategy, 
where the detector scans the CMB sky in circles which fall 
on top of each other on consecutive rotations of the instrument. 
In order to reduce the amount of data, the data from consecutive 
scan circles is averaged, a process called 'coadding'. 
We call one coadded circle a 'ring'. 
In the nominal \Planck scanning strategy, the same circle is 
scanned 60 times before repointing. 
We denote by $M$ the number of coadded circles.
On each ring, we model the correlated noise component 
as a linear combination of some simple orthogonal arithmetic 
functions, such as sine and cosine functions or Legendre 
polynomials.
In the simplest case we fit only uniform baselines.

The \Madam\ algorithm can easily be generalized to a scanning 
strategy with no coadding, by setting the coadding factor $M$ 
equal to one. We consider both types of scanning strategy in the 
simulation section. 
If no coadding is done, one can choose any length of data to 
represent a ring. The concept of ring then loses its connection 
to the spin period and its length becomes a freely chosen parameter.

The most frequently used parameters and symbols of this paper have 
been collected in Table \ref{tab:param}.

\begin{table}
\caption[a]{\protect\small Main parameters and symbols used in 
this paper and values used in simulation.}
\begin{center}
\begin{tabular}{lll}
\hline
 symbol & parameter & value \\
\hline
  $n$           &  number of samples/ring          & 4608      \\
  $N$           &  number of rings                 & 8640      \\
  $M$           &  coadding factor                 & 60        \\
  $L$           &  number of base functions        & 1-65      \\
  $\fsample$    & sampling frequency               & 76.8 Hz      \\
  $\fspin$      & spin frequency of the spacecraft & 1/60 Hz      \\
  $\sigma$      & white noise std                  & 2700 \muK    \\
  $\fmin$       & minimum freq. (noise spectrum)   & $10^{-5}$ Hz \\
  $\fknee$      & knee frequency (noise spectrum)  & 0.1 Hz       \\
  $\gamma$      & slope of the noise spectrum      & -1.0         \\ 
  $\ve m$       & pixelized CMB map                & \\
  $\bP$         & pointing matrix                  & \\
  $\bC_n$       & white noise covariance           & \\
  $\bF$         & base function matrix             & \\
  $\ve a$       & base function amplitudes         & \\
  $\tilde\ve a$ & reference values for $\ve a$     & \\
  $\bC_a$       & covariance of amplitudes $\ve a$ & \\
  $\bC^k$       & $k$th component covariance       & \\
  $b_k$         & ... and its coefficient          & \\
  $g_k$         & $k$th characteristic frequency   & \\
\hline
\end{tabular}
\end{center}
\label{tab:param}
\end{table}

The \Madam\ map-making algorithm consists of the following steps.

\begin{enumerate}
\item
Choose a set of base functions to model
the correlated noise component and compute the covariance 
matrix $\bC_a$ of their amplitudes. 
The computation of the covariance matrix is discussed 
in Section 4.

\item
Using the conjugate gradient technique, solve $\ve a$ 
from  Eq. (\ref{linc}).
The tricky part here is the evaluation of the term $\bC_a^{-1}\ve a$,
since matrix $\bC_a$ is very large. 
For \Planck-like data its dimension typically varies from
thousands to hundreds of thousands. 
Fortunately, the matrix has symmetries, 
which allow us to evaluate  $\bC_a^{-1}\ve a$ in a quite 
efficient manner, as we show in Section \ref{sec:crout}.

\item
Solve the CMB map according to Eq. (\ref{mlmap}).
This means simple binning of the destriped TOD into pixels, 
weighting by the white noise covariance.
Here we have used HEALPix
\footnote{http://www.eso.org/science/healpix}
(G\'orski et al. 1999, 2004)
pixelization.
\end{enumerate}


\subsection{Evaluation of $\bC_a^{-1}\ve{a}$}
\label{sec:crout}

Conjugate gradient solution of Eq. (\ref{linc}) requires that we 
evaluate 
\beq
    \ve{x}=\bC_a^{-1}\ve{a}  \label{xaeq}
\eeq
several times for different $\ve{a}$.
Here $\ve{a}$ and $\ve{x}$ are vectors of $(NL)$ elements, 
where $N$ is the number of rings and $L$ is the number of base 
functions per ring.
Matrix $\bC_a$ has dimension $(NL,NL)$ and is thus expensive to 
invert. However, $\bC_a$ has symmetries which allow us 
to perform the inversion quite efficiently.

We use index notation in the following.
Evaluation Eq. (\ref{xaeq}) is equivalent to solving $x_{il}$ from
\beq
   a_{il} = \sum_{i'l'}C_{a,ii'll'} x_{i'l'}. \label{caeq}
\eeq
Here $i,i'$ label rings and $l,l'$ label base functions.
The matrix has the symmetry property $C_{a,ii'll'}=C_{a,i'il'l}$.

We assume that the properties of the correlated noise component
do not change with time.
Matrix $\bC_a$ then depends on indices $i,i'$ only through 
their difference, being thus approximately circulant 
in indices $i,i'$. The matrix can be stored as a table of $L^2N$ elements.

A general symmetric matrix equation of moderate size 
can be solved by Cholesky decomposition. 
Crout's algorithm to find the decomposition
is given e.g. in Press et al. (1992).
On the other hand, circulant matrix equations can be solved 
by the Fourier technique.

We solve equation (\ref{caeq}) using a combined technique, 
where we apply Cholesky decomposition to the indices $l,l'$, 
and Fourier technique to indices $i,i'$.

We drop indices $i,i'$ for a while and introduce the following notation.
We denote by $\hat C_{ll'}$ an $(N,N)$ submatrix of $\bC_a$.
Now $\hat C$ can be understood as an $(L,L)$ matrix whose elements 
are themselves $(N,N)$ matrices.
Similarly, we denote by $\hat a_l$ an $N$-element subvector of $\ve a$.

We now have
\beq
   \hat a_l = \sum_{l'}\hat C_{ll'}\hat x_{l'}  \label{ahat}
\eeq
where it must be understood that each term 
$\hat C_{ll'}\hat x_{l'}$ involves a matrix multiplication of order $N$.
Matrix $\hat C$ has the symmetry property 
$\hat C_{ll'}=(\hat C_{l'l})^T$.
Especially, the diagonal elements $\hat C_{ll}$ are symmetric.

We apply Crout's algorithm to Eq. (\ref{ahat}).
We follow the procedure presented in Press et al. (1992), keeping in
mind that instead of scalar elements we are now 
operating with matrices. 

We want to decompose $\hat C$ into 
\beq
   \hat C_{lk}=\sum_j \hat L_{lj}\hat L_{kj}^T
\eeq
where $\hat L_{lj}=0$ for $j>l$. Here $\hat L$ is a lower triangular matrix, 
whose elements are again $(N,N)$ matrices.
Note that the transpose sign refers to the element $\hat L_{kj}$, 
not $\hat L$ itself.

We write
\bea
   \hat C_{lk} &=& \hat L_{lk}\hat L_{kk}^T
         +\sum_{j<k}\hat L_{lj}\hat L_{kj}^T  \quad(l>k)  \\
    \hat C_{ll} &=& \hat L_{ll}\hat L_{ll}^T
         +\sum_{j<l}\hat L_{lj}\hat L_{lj}^T.            
\eea
From this we can solve the elements of $\hat L$,
\bea
   \hat L_{lk} &=& [\hat L_{kk}^{-1}(\hat C_{lk}^T-\sum_{j<k}
                     \hat L_{kj}\hat L_{lj}^T)]^T  \quad (l>k)
                                                \label{llk}  \\
   \hat L_{ll} &=& [\hat C_{ll}-\sum_{j<l}\hat L_{lj}
                      \hat L_{lj}^T]^{1/2}.    \label{ldiag}
\eea
For each $l$, we first use Eq. (\ref{llk}) to solve $\hat L_{lk}$ 
for $k=1\ldots l-1$ and then Eq. (\ref{ldiag}) to solve $\hat L_{ll}$.
Once $\hat L$ is known, elements $\hat x_k$ can be solved by
 backsubstitution as
\bea
    \hat z_k &=& \hat L_{kk}^{-1}
         (\hat a_k-\sum_{j<k}\hat L_{kj}\hat z_j)
            \qquad k=1\ldots L  \label{lbks1}               \\
    \hat x_k &=& \hat L_{kk}^{-1}
         (\hat z_k-\sum_{j>k}\hat L_{jk}^T\hat x_j)
            \qquad k=L\ldots 1.   \label{lbks2}
\eea

The procedure presented above contains operations between 
$(N,N)$ matrices and vectors of length $N$. 
These $(N,N)$ matrices are nearly circulant, except for the corners,
 where they do not 'wrap around' like circulant matrices do. 
However, the deviation is small, and we may treat the matrices 
as circulant.

Circulant matrix operations are most conveniently carried out 
in Fourier space.
We pad each vector with zeros up 
to the next power of two, and use the Fast Fourier Transform
(FFT) technique to perform the Fourier transforms. 

To each elementary operation involving a circulant matrix
corresponds an operation in Fourier space. 
The corresponding
Fourier operations are the following.

\begin{enumerate}
\item Matrix multiplication 
-- element by element multiplication of Fourier transforms
\item Matrix transpose
-- complex conjugate of the Fourier transform
\item Square root of a matrix
-- Square root of the Fourier transform. 
\item Matrix inversion 
-- inverse of the Fourier transform.
\end{enumerate}

The procedure on determining $\ve{x}=\bC_a^{-1}\ve{a}$ can be 
summarized as follows.
First perform Fourier transform to the circulant ring index $i-i'$ 
of $\bC_a$. 
Then carry out Cholesky decomposition in index $l$ as described above, 
and store the resulting $\hat L$ matrix. 

For each vector $\ve{a}$, carry out a Fourier transform along 
the ring index $i$,
perform backsubstitution as given by (\ref{lbks1})-(\ref{lbks2}), 
and do an inverse Fourier transform to obtain $\ve x$.

The total operation count of the above procedure is proportional to 
$L^3N\ln{N}$, as contrasted to $L^3N^3$ of normal matrix inversion.
Furthermore, the decomposition can be done 'in place' in the space of 
$L^2N$ elements, instead of $L^2N^2$.

\subsection{Covariance of the output map}
\label{sec:outcov}

The covariance of the output map of \Madam\ is given by
\beq
    \bC_m = [\bP^T(\bC_n+\bF\bC_a\bF^T)^{-1}\bP]^{-1}, 
                                          \label{mapcov}
\eeq 
assuming that the noise is well modelled by the noise model
(\ref{noisetod}).
The inverse of (\ref{mapcov}) can be put into the form
\bea
   \lefteqn{\bC_m^{-1} =}             \label{mapcovinv}  \\
   && \bP^T\bC_n^{-1}\bP   -\bP^T\bC_n^{-1}\bF
                (\bC_a^{-1}+\bF^T\bC_n^{-1}\bF)^{-1}
                         \bF^T\bC_n^{-1}\bP. \nonumber
\eea

We presented in Section \ref{sec:crout} a procedure for evaluating
$\bC_a^{-1}\ve a$ for arbitrary $\ve a$. 
By running this procedure $L$ times 
one can compute the inverse of matrix $\bC_a$.
Matrix $\bC_a^{-1}+\bF^T\bC_n^{-1}\bF$ can then again be 
decomposed and stored using the same procedure.
When this is done, one can then easily compute any element of
 $\bC_m^{-1}$ using formula (\ref{mapcovinv}).


\section{Covariance of the component amplitudes}

\subsection{General}
\label{sec:covgen}

In this section we consider the computation of the covariance 
matrix  $\bC_a$.

First we define reference values for the amplitude vector $\ve a$.
Suppose the actual coadded noise TOD, denoted by $\ve u$, is known.
We consider here the correlated noise component only.
We fit the noise model $\bF\ve a$ to the noise stream. 
A least-squares fit gives  
\beq
   \tilde\ve a = (\bF^T\bF)^{-1}\bF^T\ve u.  \label{vecrefval}
\eeq
Eq. (\ref{vecrefval}) gives the reference values or best estimates 
for the amplitude vector. 
We compute the covariance matrix as
\beq
    \bC_a = \langle\tilde\ve a\tilde\ve a^T\rangle .
\eeq

Let now $y_p$ be the original, uncoadded noise stream.
We assume that the noise is stationary and its auto-correlation 
$c_{p-p'}=\langle y_py_{p'}\rangle$ is known.

The coadded noise stream is 
\beq
   u_{ij} = \frac1M \sum_{m=0}^{M-1} y_{[iMn+mn+j]}.
\eeq
Here $n$ is the number of samples per ring and $M$ is the number 
of coadded circles
($M=60$ for the nominal \Planck scanning strategy).
Index $i=0\cdots N-1$ labels rings, 
$j=0\cdots n-1$ labels samples on a ring, 
and $m=0\cdots M-1$ labels circles coadded into a ring. 

Let $F_{lj}$ be the values of the base functions 
$l=1\cdots L$ on a ring.
We assume orthogonality and write 
\beq
    \bar F_{lj} = \big(\sum_{j'}F^2_{lj'}\big)^{-1}F_{lj}.
\eeq
The reference values for component amplitudes are,
according to (\ref{vecrefval}), given by
\beq 
   \tilde a_{il} = \sum_{j=0}^{n-1}\bar F_{lj}u_{ij} 
         = \sum_{j=0}^{n-1}\bar F_{lj}
              \frac1M \sum_{m=0}^{M-1} y_{[iMn+mn+j]}. 
                                        \label{refval}  
\eeq
For uniform baselines, for instance, the reference value is 
simply the average of the noise over the ring.

Next we calculate the theoretical covariance 
of the reference values (\ref{refval}).
The elements of the covariance matrix are given by
\bea
   \lefteqn{C_{a,ii'll'} = 
        \langle \tilde a_{il}\tilde a_{i'l'}\rangle} \label{gencov1} \\
    &&= \sum_{j,j'=0}^{n-1}\bar F_{lj}\bar F_{l'j'} 
           \frac{1}{M^2}\!\!\sum_{m,m'=0}^{M-1}    
       \langle y_{[inM+mn+j]}y_{[i'nM+m'n+j']}\rangle   \nonumber \\
   &&= \sum_{j,j'=0}^{n-1}\bar F_{lj}\bar F_{l'j'} 
     \frac{1}{M^2}\!\!\sum_{m,m'=0}^{M-1}    
        c_{[(i-i')nM+(m-m')n+j-j']}.     \nonumber        
\eea
The sum over $m,m'$ can be combined into one sum over $m''=m-m'$,
\bea
  \lefteqn{C_{a,ii'll'}=}  \label{gencov2}  \\
      && \sum_{j,j'=0}^{n-1}\bar F_{lj}\bar F_{l'j'}
           \!\!\!\sum_{m''=-M}^M\frac{M-|m''|}{M^2}  
            c_{[(i-i')nM+m''n+j-j']}.  \nonumber     
\eea
This is a general formula for the elements of the covariance 
matrix $\bC_a$.


\subsection{Exponential expansion of the auto-correlation function}
\label{sec:covexpansion}

Covariance (\ref{gencov2}) is rather heavy to evaluate 
computationally, since it contains a three-dimensional sum.
However, if the auto-correlation function can be expanded 
in exponential functions, the covariance can be computed in a very
efficient way. This holds, e.g., for $1/f$ and $1/f^2$ noise.

Suppose the auto-correlation function can be expanded
as 
\beq
   c(t) = \sum_k b_kc_k(t)  \label{eexpansion}
\eeq
where
\beq
   c_k(t) = \exp(-g_k|t|)  \label{autok}
\eeq
where $g_k$ is a selected set of characteristic frequencies and
and $b_k$ are coefficients to be determined.

We denote by $\bC^k$ the covariance matrix that corresponds
to an exponential auto-correlation function of the form (\ref{autok}).
Once the component covariances $\bC^k$ and coefficients $b_k$
have been determined, the total covariance $\bC_a$ can be 
computed as 
\beq
   \bC_a = \sum_kb_k\bC^k.  \label{caexpansion}
\eeq
The component covariances $\bC^k$ can be computed in a very efficient 
way making use of the basic properties of the exponent function.

Expanding the auto-correlation as (\ref{eexpansion}) is equivalent to 
expanding the power spectrum as
\beq
    P(f) = \sum_kb_k\frac{2g_k}{g_k^2+(2\pi f)^2}. \label{powspecsum} 
\eeq
The expansion does not exist for arbitrary noise spectra.
In Appendix A we calculate the 
expansion explicitly for a power-law spectrum  of the form
\beq
     P(f) = \left(\frac{\fknee}{f}\right)^\gamma
           \frac{\sigma^2}{\fsample}
          \quad (f>\fmin).
\eeq
for $-2\le\gamma<0$.
Here $\fknee$ is the knee frequency, at which the spectral power 
equals the white noise power $\sigma^2/\fsample$, $\sigma$ is the 
white noise std, and $\fsample$ is the sampling frequency.

For $1/f$ noise ($\gamma=-1$) the expansion is particularly simple.
If the characteristic 
frequencies $g_k$ are chosen uniformly in logarithmic scale,
the correct spectrum is obtained with
\beq
     b_k = 2\sigma^2\frac{\fknee}{\fsample} \Delta,  \label{bkoof} 
\eeq
where $\Delta$ is the logarithmic interval in $g_k$. 

\begin{figure}
\center{\resizebox{\hsize}{!}{\includegraphics{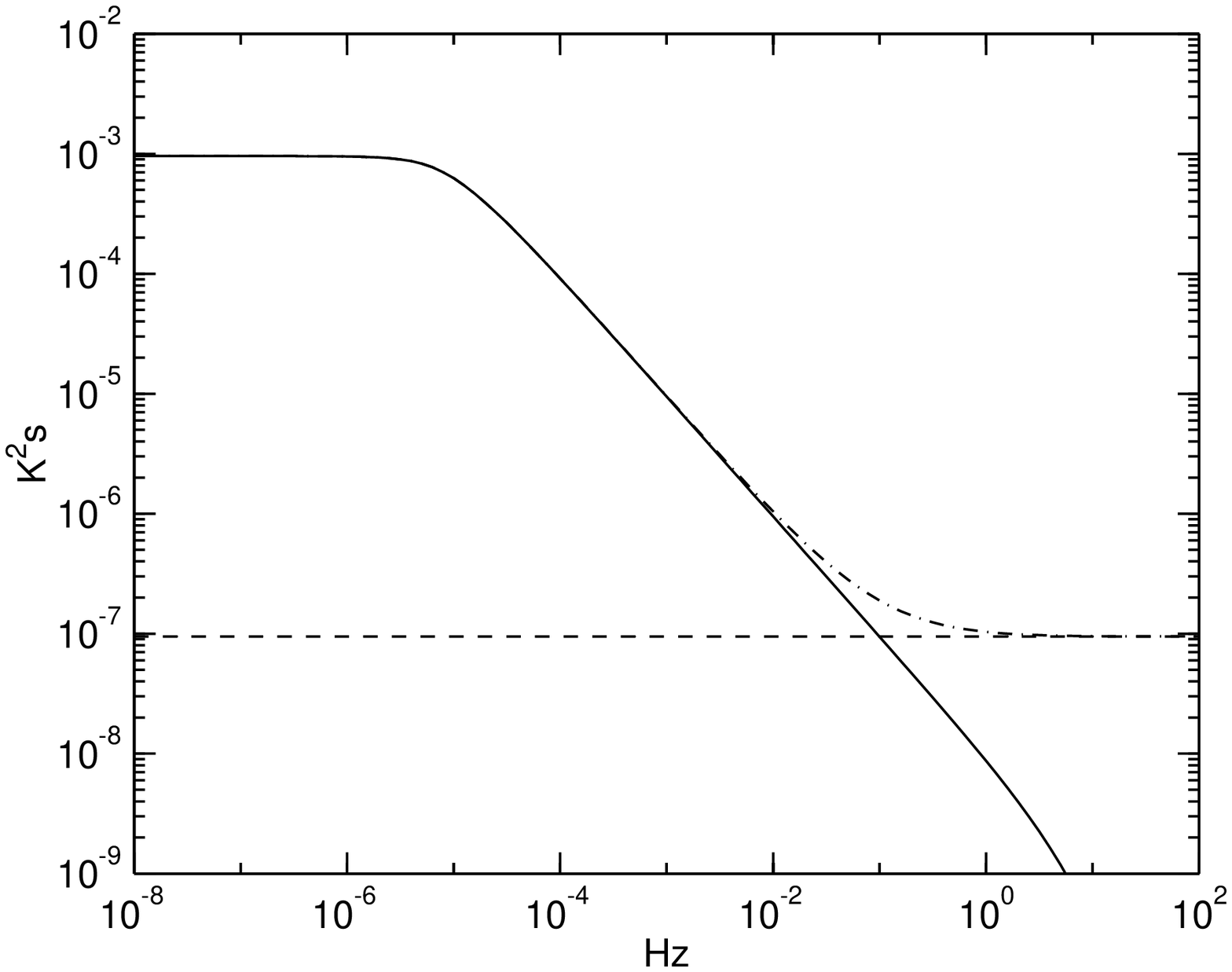}}}
\caption{$1/f$ noise power spectrum, as given by expansion
(\ref{powspecsum}) and Eq. (\ref{bkoof}). 
The solid line shows the pure 
$1/f$ noise spectrum. 
The dashed line shows the white noise level 
($P_{\rm wn}=\sigma^2/\fsample$).
The dash-dotted line shows the total noise spectrum, including 
both components. 
The parameters were $\sigma=2700$ \muK, $\fsample=76.8$ Hz, $\fknee=0.1$ Hz, 
$\Delta=1$, $\fmin=10^{-5}$ Hz, $\fmax=10$ Hz.} 
\label{fig:powspec}
\end{figure}

The $1/f$ spectrum, as given by the expansion (\ref{powspecsum}) 
with coefficients (\ref{bkoof}),
is shown in Fig. \ref{fig:powspec}.
The $1/f$ form holds inside the frequency range 
$\fmin<f<\fmax$
spanned by the characteristic frequencies $g_k$.
Below $\fmin$ the spectrum levels out, as can be seen from 
the figure. 

Another simple example is the $1/f^2$ spectrum ($\gamma=-2$). 
In that case the desired spectrum is given by one single $g$ 
component with coefficient
\beq
     b = \frac{2\pi^2\fknee^2}{g}\frac{\sigma^2}{\fsample}
\eeq
and $g=2\pi\fmin$.

\subsection{Component covariance matrices}

In this section we give explicit formulae for the elements of 
the component covariance 
matrix $\bC^k$ that corresponds to the auto-correlation function 
(\ref{autok}). Derivation is given in Appendix A. 
Here we just quote the results.

We use again the index notation, where indices $i,i'$ refer to rings
and $l,l'$ to base functions. The elements of the component covariance 
matrix are given by
\beq
   C^k_{ii'll'} =                  
      G_k S^+_{kl} S^-_{kl'}\exp\big(-\frac{g_k}{\fspin}(i-i'-1)M\big) 
                      \quad(i>i')  \label{covpos}  
\eeq
\beq
   C^k_{ii'll'} =                   
      G_k S^-_{kl} S^+_{kl'}\exp\big(-\frac{g_k}{\fspin}(i'-i-1)M\big) 
                      \quad(i<i')   \label{covneg}  
\eeq
\bea
   \lefteqn{C^k_{ii'll'} = \frac1M \sum_{j=0}^{n-1}
       \big(\bar F_{lj}\bar F_{l'j}
       + \bar F_{lj}A_{l'j}+\bar F_{l'j}A_{lj}\big)}  \label{covzero} \\    
       &&\hspace{10mm}\qquad  +G_k^0(S^+_{kl}S^-_{kl'}+S^-_{kl}S^+_{kl'})  .
                                     \qquad (i=i')    \nonumber 
\eea
Here $\fspin=\fsample/n$, where $n$ is the number of samples on a ring,
represents the 
spin frequency of the instrument. If no coadding is applied,
$n$ can be chosen freely, and $\fspin$ does not need to have 
any connection to the scanning pattern of the instrument. 
In that case $\fspin$ represents simply the inverse of 
the chosen baseline length.

Factors $S^+$ and $S^-$ are defined as
\beq
   S^+_{kl} = \sum_{j=0}^{n-1}\bar F_{lj}
              \exp(-\frac{g_k}{\fsample}j)\label{splus}
\eeq
\beq
   S^-_{kl} = \sum_{j=0}^{n-1}\bar F_{lj}
              \exp(-\frac{g_k}{f_{s}}(n-j)).  \label{sminus}
\eeq

Coadding brings in the factors
\beq
    G_k = \frac{1}{M^2} \frac{[1-\exp(-\frac{g_k}{\fspin}M)]^2}
            {[1-\exp(-\frac{g_k}{\fspin})]^2} 
\eeq
and
\beq
     G_k^0 = \frac1M\frac{1}{1-\exp(-\frac{g_k}{\fspin})}
     \Big[ 1-\frac1M\frac{1-\exp(-\frac{g_k}{\fspin}M)}
                         {1-\exp(-\frac{g_k}{\fspin})} 
                                       \Big]  
\eeq
If no coadding is done ($M=1$), we have $G_k=1$ and $G_k^0=0$. 

Factor $A_{lj}$ is defined by
\beq
     A_{lj} = \sum_{j'=0}^{j-1}\bar F_{lj'} 
               \exp(-\frac{g_k}{\fsample}(j-j'))  \label{asum}
\eeq
and can be computed rapidly using the recurrence relation
\beq
      A_{lj} =( A_{l,j-1}+\bar F_{l,{j-1}}) 
          \exp(-\frac{g_k}{\fsample}),  \label{recurr}
\eeq
with the starting value $A_{l0}=0$.

Formulae (\ref{covpos})-(\ref{covzero}) are very fast to evaluate 
numerically, as compared to the general formula (\ref{gencov2}).

\begin{figure}
\center{\resizebox{\hsize}{!}{\includegraphics{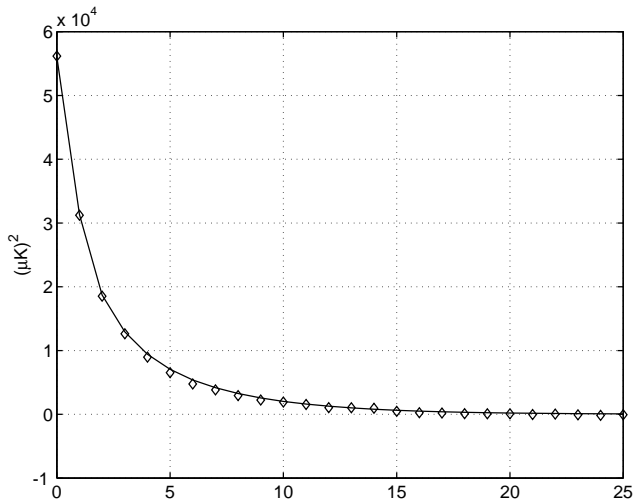}}}
\caption{Covariance between uniform baselines, as a function of 
the distance between rings.
The solid line shows the theoretical curve, computed using 
Eqs. (\ref{covpos})-(\ref{covzero}). Diamonds present the
values determined from simulated noise.
The noise parameters were $\sigma=2700$ \muK, $\fsample=76.8$ Hz, 
$\fknee=0.1$ Hz, 
$\Delta=1$, $\fmin=10^{-5}$ Hz, $\fmax=10$ Hz.}  
\label{fig:noisecov}
\end{figure}


\begin{table}
\caption[a]{\protect\small The first elements of the covariance
 matrix $\bC_a$ (in ({\muK})$^2$). The noise parameters were the same 
 as in Fig. \ref{fig:noisecov}. We used Fourier components as base 
 functions, normalized as $\sum_jF_{lj}^2=n$.
 We show elements $l,l'=1\ldots5$ and $i-i'=0\ldots3$.
 Index value $l=1$ refers to the uniform baseline and 
 values $l=2$ and $l=3$ ($l=4$ and $l=5$) to the sine and cosine 
 of the first (second) Fourier mode, respectively.
 The noise parameters were the same as in Fig. \ref{fig:noisecov}.
  }
\begin{center}
\begin{tabular}{llrrrrr}
\hline
$l$ & i-i' & $l'=1$ & $l'=2$ & $l'=3$ & $l'=4$ & $l'=5$ \\
\hline
1 & 0 &  56049 &   0     &  -2.41    &    0    &  -0.668    \\
  & 1 &  31324 &  -89.6  &   1.10    &  -44.9  &   0.307    \\
  & 2 &  18707 &  -28.9  &   0.0531  &  -14.5  &   0.0133   \\
  & 3 &  12928 &  -16.3  &   0.0210  &   -8.17 &   5.25e-3  \\
 \hline
2 & 0 &   0    &   161  &   0        &   1.64    &    0       \\
  & 1 &  89.6  &  -1.42 &   0.209    &  -0.745   &  0.0696    \\
  & 2 &  28.9  &  -0.0751 &   2.35e-4 &  -0.0376  &  5.88e-5  \\
  & 3 &  16.3  &  -0.0297 &   5.90e-5 &  -0.0148  &  1.48e-5  \\
\hline
3 & 0 & -2.41    &  0        &   158      &   0         & -0.123    \\
  & 1 &  1.10    & -0.209    &   0.133   &  -0.139     &  0.0617   \\
  & 2 &  0.0531  & -2.35e-4 &   1.14e-6 &  -1.18e-4  &  2.85e-7 \\
  & 3 &  0.0210  & -5.90e-5 &   1.73e-7 &  -2.95e-5  &  4.32e-8 \\
\hline
4 & 0 &     0   &  1.64    &   0    &   79.9   &  0        \\
  & 1 &  44.9   & -0.745   &  0.139 &  -0.401   &  0.0524   \\
  & 2 &  14.5   & -0.0376  &  1.18e-4  &  -0.0188  &  2.94e-5 \\
  & 3 &   8.17  & -0.0148  &  2.95e-5  &  -7.42e-3 &  7.38e-6 \\
\hline
5 & 0 & -0.668    &  0        &  -0.123    &   0        &   79.0     \\
  & 1 &  0.307    & -0.0696   &   0.0617   &  -0.0524   &   0.0334   \\
  & 2 &  0.0133   & -5.88e-5 &   2.85e-7 &  -2.94e-5 &   7.13e-8 \\
  & 3 &  5.25e-3  & -1.48e-5 &   4.32e-8 &  -7.38e-6 &   1.08e-8 \\
\hline
\end{tabular}
\end{center}
\label{tab:covariance}
\end{table}

Figure \ref{fig:noisecov} presents the theoretical covariance,
computed using expansion (\ref{caexpansion}), between 
uniform baselines for $1/f$ noise with $\fknee=0.1$ Hz. 
Other parameters used were $\fsample=76.8$ Hz, $\fspin=1/60$ Hz,
$n=4608$ and $M=60$. 
We show in the same figure
the covariance as determined from simulated $1/f$ noise.
We generated 10 realizations of noise TOD of one year length,
and computed their auto-correlation using the Fourier technique.
The agreement is very good. 

As another example we show in Table \ref{tab:covariance} 
the first elements of the
covariance matrix for Fourier components.
Index $l=1$ refers to the uniform baseline and 
indices $l=2$ and $l=3$ ($l=4$ and $l=5$) to the sine and cosine 
of the first (second) Fourier mode, respectively.
We have normalized all components to $\sum_jF_{lj}^2=n$. 
Elements of matrix $\bF$ are thus
$F_{1j}=1$, 
$F_{2j}=\sqrt2\sin(2\pi j/n)$,
$F_{3j}=\sqrt2\cos(2\pi j/n)$,
$F_{4j}=\sqrt2\sin(4\pi j/n)$,
$F_{5j}=\sqrt2\cos(4\pi j/n)$,
and $\bar\bF$ is given by $\bar F_{lj}=F_{lj}/n$.
We see that the dominant elements are those corresponding 
to uniform baselines.


\section{Simulation results}
\label{sec:simu}

\subsection{Data sets}
\label{sec:datasets}

We have produced two sets of simulated TOD.
We refer to them as 'coadded' and 'uncoadded' data sets.

The coadded data set mimicks the one year TOD from one 
\Planck LFI 70 GHz detector.
The scanning pattern was the following. The spin axis remained in the 
equatorial plane and was turned 2.4 arcmin every hour,
so that after 8640 hours the spin axis had turned 360 degrees. 
The detector turned around the spin axis with an opening angle of 
85 $\deg$ and spin frequency $\fspin=1/60$ Hz. 
The sampling frequency was $\fsample=76.8$ Hz.
We coadded data of 60 consecutive spin circles to form a ring.
Our total data set consisted of 8640 rings, with 4608 samples 
in each. The sky coverage was 0.9964.

The uncoadded data set was generated with a quite similar 
scanning pattern. The main difference was that instead of 
moving in steps, the spin axis
turned continuously at the rate of 360 degrees in 8640 min. 
The sampling and spin frequencies as well as the opening angle 
were the same as in the first data set.
Because the spin axis moved continuously, consecutive circles 
did not fall on top of each other, and no coadding was done.
The total length of the TOD was the same as in the coadded
data set, i.e. 8640$\times$4608 samples.
The amount of data was equivalent to 6 days of one detector 
\Planck data, spread over the whole sky. 
This scanning pattern was rather artificial, but our purpose was
only to demonstrate the use of \Madam\ in the case of uncoadded 
data. Full-scale simulations with realistic uncoadded one-year 
\Planck data are beyond the scope of this paper.

The underlying CMB map was created by the Synfast code of the 
HEALPix package (G\'orski et al. 1999, 2004), 
starting from the CMB anisotropy angular power spectrum computed 
with the CMBFAST\footnote{http://www.cmbfast.org} code 
(Zeljak \& Zaldarriaga 1996)
using the cosmological parameters $\Omega_{\rm tot}=1.00$,
$\Omega_\Lambda=0.7$, $\Omega_\mathrm{b}h^2=0.02$, $h=0.7$, $n=1.00$, 
and $\tau=0.0$. 
We created the input map with HEALPix resolution
$N_\mathrm{side}=2048$ and with a symmetric Gaussian beam with 
full width at half maximum (FWHM) of 14 arcmin. We then formed 
the signal TOD by picking temperatures from this map. 
Our output maps have resolution parameter
$N_\mathrm{side}$=512, corresponding to an angular resolution of 
7 arcmin.

We used the Stochastic Differential Equation (SDE) technique to
create the instrument noise stream, which we added to the signal
TOD. We generated noise with
power spectrum \beq
   P(f) = \left(1+\frac{\fknee}{f}\right)\frac{\sigma^2}{\fsample},
          \quad(f>\fmin)
\eeq with parameters $\sigma=2700$ {\muK}, knee frequency $\fknee=0.1$ Hz, and
$\fmin=10^{-5}$ Hz. 
The white noise level 2700 \muK\ (CMB temperature
scale) corresponds
to the estimated white noise level of one 70 GHz LFI
detector.
We used the same noise spectrum for both data sets.

We run our code on one processor of an IBM eServer Cluster 
1600 supercomputer.

\subsection{Results for coadded data}
\label{sec:coadded}

We show our results for the coadded data set in Tables
 \ref{tab:rmsc}-\ref{tab:signal}.
As a figure of merit we have used the rms of the residual noise map.
The residual noise map was computed by subtracting from the output 
map a reference map. The reference map was created by coadding the
pure signal TOD into a map. We then subtracted the monopole from
the residual map and computed its rms value.


\begin{table}
\caption[a]{\protect\small Average rms of the
residual noise map for different numbers of base 
functions, for coadded data. We fit Fourier components 
and Legendre polynomials. 
The averages are taken over 10 noise realizations.
We show also the number of iteration steps and CPU time 
in the Fourier case.}
\begin{center}
\begin{tabular}{rllrr}
\hline
           & Legendre   & Fourier            \\
\hline
$L$        & rms$/\mu$K & rms$/\mu$K & iter & CPU/s\\
\hline
1      &  110.634  &  110.634  & 16 & 59   \\ 
2      &  110.560  &     \\
3      &  110.525  &  110.485  & 20 & 65   \\
4      &  110.481  &     \\
5      &  110.451  &  110.422  & 28 & 106  \\
7      &  110.413  &  110.386  & 28 & 123  \\
9      &  110.387  &  110.363  & 32 & 164  \\
11     &  110.368  &  110.347  & 32 & 191  \\
15     &  110.344  &  110.326  & 36 & 267  \\
25     &  110.312  &  110.301  & 40 & 530  \\
35     &  110.298  &  110.290  & 44 & 758  \\
45     &  110.290  &  110.283  & 52 & 1154 \\
65     &           &  110.277  & 56 & 1959 \\
\hline
\end{tabular}
\end{center}
\label{tab:rmsc}
\end{table}

\begin{figure}
\center{\resizebox{\hsize}{!}{\includegraphics{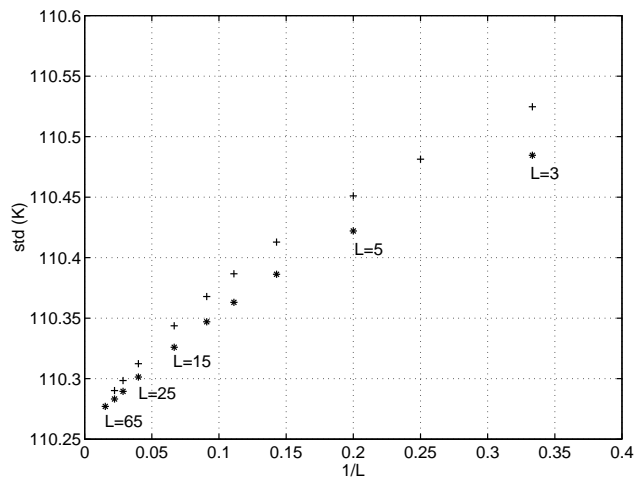}}}
\caption{Rms of the residual noise map plotted against the inverse 
of the number of base functions, for Legendre polynomials (+) 
and Fourier components (*). The numerical values are given in 
Table \ref{tab:rmsc}.
The std seems to converge towards value 110.26 \muK\ at the limit 
$L\rightarrow n$.} 
\label{fig:stdconvergence}
\end{figure}

The results for different numbers of base functions are given
in Table \ref{tab:rmsc}. 
The given rms values are averages over 10 noise realizations.
We tried two sets of base functions:
Fourier components and Legendre polynomials.
Because a Fourier fit always includes an uniform baseline plus 
an equal number of sine and cosine functions, 
the total number $L$ of base functions is always an odd number.
The rms values continue improving when we increase the number 
of base functions.
Fourier components give lower rms values than Legendre polynomials
for the same number of components. In the rest of the
simulations in this section we fitted only Fourier components.

We show also the number of iterations and total CPU time taken by one 
run in the case of Fourier components. Since the CPU time naturally 
depends on the computer used and may vary from run to run, 
the values quoted 
should not be taken too seriously. They merely give an idea how 
the run time increases with increasing number of base functions. 

In Fig. \ref{fig:stdconvergence} we have plotted the rms values
against the inverse of the number of base functions.
At the limit $1/L\rightarrow 0$ the rms values seem to be 
approaching the value 110.26 \muK. We expect that to be the std of 
the minimum-variance map (Section \ref{sec:gls}).
The expected contribution from white noise to the residual 
map rms is 108.95 \muK. This value was computed from the white noise
sigma and the known distribution of measurements in the sky.

We have compared results of fitting uniform baselines using \Madam\ 
and ordinary destriping technique.
We got the destriping results by running \Madam\ with $\bC_a^{-1}=0$.
At this limit the method reduces to pure destriping.
The destriping result was 110.63444 \muK\ (110.63443 \muK\ with \Madam).
This indicates that when fitting uniform baselines only, 
the covariance plays very little role, but the baselines can be
determined from the data alone with good accuracy.

Keih\"anen et al. (2004) showed that fitting Fourier components 
beyond the uniform baseline with the ordinary destriping technique,
without using the covariance matrix,
did not improve the results. In this work we have found a clear
improvement. This indicates, that information about the noise 
spectrum  is important when fitting base functions other than 
the uniform baseline.


\begin{table}
\caption[a]{\protect\small 
Effect of misestimating the noise spectrum.
We show the average residual map std (in \muK) for different 
numbers of Fourier components and for different assumed
knee frequencies $\fknee$ (upper panel), spectral slopes $\gamma$ 
(middle), and minimum frequencies $\fmin$ (lower panel).
The other two noise parameters were kept at their correct values
($\fknee=0.1$ Hz, $\gamma=-1.0$, and 
$\fmin=10^{-5}$ Hz). 
The lowest rms value on each row is denoted by an asterisk.
The correct parameter value is shown in boldface.}
\begin{center}
\begin{tabular}{rllll}
\hline
 & & $\fknee$ \\
L & 0.03 Hz & 0.05 Hz & \bf0.1 Hz & 0.2 Hz \\
\hline
1  & 110.634* & 110.634* &  110.634* &  110.634*  \\ 
3  & 110.492 &  110.484* &  110.485  &  110.492  \\
5  & 110.436 &  110.420* &  110.422  &  110.441  \\
7  & 110.409 &  110.387 &  110.386*  &  110.411  \\
9  & 110.392 &  110.366 &  110.363*  &  110.393  \\
11 & 110.382 &  110.353 &  110.347*  &  110.380  \\
15 & 110.369 &  110.335 &  110.326*  &  110.363  \\
25 & 110.355 &  110.319 &  110.301*  &  110.344  \\
\hline
 & & $\gamma$  \\
L & -0.6 & -0.8 & \bf -1.0 & -1.2 \\
\hline
1   &  110.640 &  110.635 &  110.634*  &  110.634  \\ 
3   &  110.490 &  110.483* &  110.485  &  110.488  \\
5   &  110.428 &  110.419* &  110.422  &  110.428  \\
7   &  110.394 &  110.384* &  110.386  &  110.393  \\
9   &  110.373 &  110.361* &  110.363  &  110.369  \\
11  &  110.359 &  110.346* &  110.347  &  110.352  \\
15  &  110.340 &  110.326* &  110.326*  &  110.330  \\
25  &  110.321 &  110.303 &  110.301*  &  110.304  \\
35  &  110.314 &  110.293 &  110.290*  &  110.292  \\
\hline
 & & $\fmin$ \\
L & $10^{-6}$ Hz & \bf$\mathbf{10^{-5}}$ Hz & $10^{-4}$ Hz &
 $10^{-3}$ Hz \\
\hline
1   &  110.634* &  110.634* &  110.635 & 110.655 \\ 
3   &  110.485* &  110.485* &  110.485 & 110.510 \\
5   &  110.422* &  110.422* &  110.422 & 110.463 \\
7   &  110.386* &  110.386* &  110.386 & 110.433 \\
9   &  110.363* &  110.363* &  110.363 & 110.415 \\
11  &  110.347* &  110.347* &  110.347 & 110.402 \\
15  &  110.326* &  110.326* &  110.326 & 110.386 \\
25  &  110.301* &  110.301* &  110.302 & 110.367 \\
\hline
\end{tabular}
\end{center}
\label{tab:misest}
\end{table}

 We have also studied the effect of misestimating the noise spectrum.
We varied in turn each of the three noise parameters (knee frequency, 
spectral slope, and minimum frequency) while keeping the other two at 
their correct values ($\fknee=0.1$ Hz, $\gamma=-1.0$, $\fmin=10^{-5}$ Hz). 
We then recomputed the covariance matrix $\bC_a$
with the new parameter values and rerun the map estimation.
We fitted Fourier components only.
The results are shown in Table \ref{tab:misest}.
 
It is perhaps surprising that for small $L$ underestimating the knee
frequency or assuming a less deep slope seems to improve the results.
This can be understood as follows. When $L$ is small, the noise is 
not perfectly modelled by the base functions. There is an error 
involved, related to the higher Fourier components that have not 
been included in the analysis. This error affects the estimation 
of the lower components, leading to a too high variation in their
amplitudes. 
The error in the covariance matrix, caused by misestimation of the 
noise spectrum, partly compensates for this error. 
We notice that the best results 
are obtained with a spectrum (less deep a slope or lower knee 
frequency) which leads to a smaller covariance for the low-frequency
Fourier components. Smaller covariance tends to restrict the variation 
of the amplitudes, thus decreasing their error also. 
With larger $L$ the phenomenon  disappears, and the lowest rms is 
obtained with the correct noise spectrum, as expected.


\begin{table}
\caption[a]{\protect\small Pixelization noise.
Rms of the residual map for noise-free TOD.
The error is due to finite pixelization of the sky.}
\begin{center}
\begin{tabular}{rl}
\hline
$L$       & rms/$\mu$K  \\
\hline
 1 &  0.138  \\
 3 &  0.207  \\
 5 &  0.233  \\
 7 &  0.257  \\
 9 &  0.274  \\
11 &  0.289  \\
15 &  0.319  \\
25 &  0.359  \\
35 &  0.380  \\
45 &  0.393  \\
\hline
\end{tabular}
\end{center}
\label{tab:signal}
\end{table}

Table \ref{tab:signal} shows results from a run with noise-free data.
The TOD contained only the contribution from the CMB signal, 
but no instrument noise. The error that still remains in the map 
is due to 'pixelization noise' (Dor\'e et al. 2001)
caused by the finite size of sky pixels. The pixelization error 
increases with
increasing number of base functions, but is very small compared with 
the error due to instrument noise.

\subsection{Results for uncoadded data}
\label{sec:uncoadded}

If no coadding is applied, the length $n$ of a ring is not determined 
by the scanning pattern of the instrument, but is a free parameter 
to be chosen at will. We then have two parameters to select: 
the number $L$ of base functions and their length $n$.

To keep things simple, we tried two schemes. First we kept 
the baseline length fixed at one minute
and varied the number 
of Fourier components that we fitted. Secondly, we fitted 
uniform baselines only ($L=1$) but varied their length.

Results from the first case are shown in Table \ref{tab:rmsu1}.
The baseline length was fixed at $n=4608$ samples (one minute).
We show again the average rms of the residual noise map, 
averaged over 10 realizations of noise. The white noise level 
is higher than in the coadded case by a factor of  $\sqrt{60}$. 
The expected white noise contribution to the map rms is 
844.0 \muK.

Table \ref{tab:rmsu2} shows results of fitting uniform 
baselines of different lengths. The first column gives 
the length $n$ of the baseline, as the number of samples.
The second column gives the baseline length in seconds.
The third column shows the number of baselines per minute 
($4608/n$). 
The shortest baseline we tried consisted of only 
9 samples. 

The third column of Table \ref{tab:rmsu2}
and the first column of Table \ref{tab:rmsu1} are 
comparable, since they give the total number of unknows 
per one minute of TOD. We see that for a given number of 
unknowns, fitting Fourier components works better than 
fitting uniform baselines. However, when we compare CPU times, 
we see that fitting uniform baselines is more effective.


\begin{table}
\caption[a]{\protect\small 
Average residual noise rms for different numbers 
of Fourier components, for the uncoadded data set.
The ring length was $n=4608$.
Also shown are the number of iteration steps and the CPU time 
for one run.}
\begin{center}
\begin{tabular}{rlrr}
\hline
$L$ & rms/$\mu$K & iter & CPU/s\\
\hline
  1 &  857.131  &  28  & 68  \\
  3 &  855.782  &  36  & 102 \\
  5 &  855.294  &  48  & 161 \\
  7 &  838.373  &  52  & 205 \\
  9 &  854.842  &  56  & 280 \\
 11 &  854.718  &  60  & 336 \\
 15 &  854.555  &  64  & 436 \\
 25 &  854.357  &  76  & 847 \\
\hline
\end{tabular}
\end{center}
\label{tab:rmsu1}
\end{table}


\begin{table}
\caption[a]{\protect\small Results of fitting uniform 
baselines of different lengths to uncoadded data. 
The first three columns give the baseline length 
as the number of samples and in seconds, respectively,
and the number of baselines per one minute of TOD.
The next columns give the average residual noise rms,
number of iteration steps and total CPU time taken by one run.}
\begin{center}
\begin{tabular}{rrrlrrl}
\hline
$n$ & $t$/s & N/min & rms/$\mu$K & iter & CPU/s \\
\hline
 4608  & 60.0  &   1  & 857.131 & 28  & 68   \\
 2304  & 30.0  &   2  & 856.220 & 32  & 80   \\
 1152  & 15.0  &   4  & 855.837 & 36  & 96   \\
  576  &  7.5  &   8  & 855.202 & 48  & 123  \\
  288  &  3.75 &  16  & 854.769 & 64  & 150  \\
  144  &  1.88 &  32  & 854.463 & 84  & 215  \\
   72  &  0.94 &  64  & 854.271 & 116 & 331  \\
   36  &  0.47 & 128  & 854.169 & 160 & 672  \\
   18  &  0.23 & 256  & 854.116 & 224 & 1284 \\
    9  &  0.12 & 512  & 854.092 & 320 & 2889 \\
\hline
\end{tabular}
\end{center}
\label{tab:rmsu2}
\end{table}


\begin{table}
\caption[a]{\protect\small Comparison between \Madam\ and destriping.}
\begin{center}
\begin{tabular}{rrrll}
\hline
    &       &       & \Madam      & destriping \\
$n$ & $t$/s & N/min & rms/$\mu$K & rms/$\mu$K \\
\hline
 4608  & 60.0  &  1  & 857.131 & 857.135 \\
 2304  & 30.0  &  2  & 856.220 & 856.239 \\
 1152  & 15.0  &  4  & 855.837 & 856.779 \\
  576  &  7.5  &  8  & 855.202 & 861.118 \\
  288  &  3.75 & 16  & 854.769 & 875.798 \\
\hline
\end{tabular}
\end{center}
\label{tab:compdestr}
\end{table}

As in the case of coadded data, we compared results of fitting 
uniform baselines using \Madam\ and ordinary destriping technique.
The results are shown in Table \ref{tab:compdestr}. 
With one minute baselines the difference between the methods is small,
but increases with decreasing $n$, in favour of \Madam. 
The rms value obtained with destriping reaches a minimum around 
0.5 min baseline length, while with \Madam\ the values continue 
improving. With small values of $n$ \Madam\ is clearly superior 
to basic destriping.


\section{Conclusions}
\label{sec:conclu}

We have presented a new map-making method for CMB experiments 
called \Madam.
The method is based on the well known destriping technique,
but unlike basic destriping, it also uses information on the
known statistical properties of the instrument noise in the form 
of the covariance matrix of the base function amplitudes.
We have shown that with this extra information the CMB map
can be estimated with better accuracy than with pure destriping.

We have tested the method with simulated coadded \Planck-like data.
As a figure of merit we have used the rms of the residual noise map.
Our simulations show that fitting more base functions clearly 
improves the accuracy of the output map, with the cost of 
increasing requirements for CPU time and memory. 

We have shown theoretically that the map estimate given by \Madam\ 
approaches the optimal minimum-variance map when the number of fitted 
base functions increases. In practice it is not possible to 
reach the exact minimum-variance map using \Madam, due to CPU time 
and memory limitations. Still, \Madam\ provides a fast and efficient
 map-making method.
By varying the number of base functions the user may flexibly 
move from a very fast but less accurate map-making (small $L$) 
to a more accurate but more time-consuming map-making (large $L$),
depending on what is desired.

We also demonstrated the use of \Madam\ for uncoadded data. 
Although the data set used was quite artificial, in the sense 
that it does not mimick data from any existing CMB experiment,
the method was shown to work well for uncoadded data also.

The current implementation of the method is a serial one.
With a parallelized version we expect to be able to process 
data sets equivalent to full-year uncoadded \Planck data.

\section*{Acknowledgments}
This work was supported by the Academy of Finland grant 75065.
TP wishes to thank the V\"ais\"al\"a Foundation for financial support.
We thank CSC (Finland) for computational resources. 
We acknowledge the use of the HEALPix package (G\'orski et al., 1999, 2004)
and CMBfast (Seljak \& Zaldarriaga, 1996).
We thank M. Reinecke for useful communication concerning $1/f^2$ noise.
The work reported in this paper was done by the CTP Working
Group of the \Planck Consortia. \Planck is a mission of the European Space
Agency.




\appendix

\section{Covariance matrix for a power-law power spectrum}

We discussed the computation of the covariance matrix $\bC_a$ 
in Section 4. In this appendix we present some technical calculations 
which were omitted there.

\subsection{Exponential expansion for a power-law spectrum}

Assume that the auto-correlation function of the noise 
can be expanded as 
\beq
   c(t) = \sum_k b_k\exp(-g_k|t|)  \label{aeexpansion}
\eeq
where
and $g_k$ is a selected set of characteristic frequencies.
We now derive the coefficients $b_k$ in the case of a power-law 
power spectrum of the form 
\beq
    P(f)\propto f^\gamma
\eeq
for $-2\le\gamma<0$.

The power spectrum $P(f)$ and the auto-correlation function $c(t)$
of stationary noise are related by the cosine transform
\beq
     P(f) = \int_{-\infty}^\infty c(t)\cos(2\pi ft)dt.
\eeq
The auto-correlation function $\exp(-g|t|)$ corresponds 
to the power spectrum
\beq
     P(f,g) = \frac{2g}{g^2+(2\pi f)^2}. \label{aoneprocess}
\eeq
The total power spectrum corresponding to the auto-correlation 
(\ref{aeexpansion}) is given by
\beq
    P(f) = \sum_kb_kP(f,g_k) 
         = \sum_k\frac{2b_kg_k}{g_k^2+(2\pi f)^2}.  \label{apowspecsum} 
\eeq

We pick the frequencies $g_k$ uniformly in logarithmic scale
inside some range $[\fmin,\fmax]$ and use the ansatz
\beq 
   b_k = Ag_k^{\gamma+1}\Delta
\eeq
where $\Delta=\ln(g_{k+1}/g_k)$ is the logarithmic step in $g$ 
and $A$ is a constant to be determined.
The total power spectrum becomes
\beq
    P(f) = A\sum_k\frac{2g_k^{\gamma+2}}{g_k^2+(2\pi f)^2}\Delta. 
\eeq
We transform the sum into an integral
\bea
    P(f) &=&  A\int_{\fmin}^{\fmax}
              \frac{2g^{\gamma+2}}{g^2+(2\pi f)^2} \frac{dg}{g}  \\
      &\approx&  A\int_0^\infty
            \frac{2g^{\gamma+1}}{g^2+(2\pi f)^2} dg           
               \quad(\fmin\ll f\ll \fmax).   \nonumber
\eea
The integral converges for $-2<\gamma<0$,
\beq 
      P(f) =  A\frac{\pi(2\pi f)^\gamma}{\sin[(\gamma+2)\pi/2]}    
             \quad\propto\,  f^{\gamma} .
\eeq
We choose
\beq
   A = \frac{\sigma^2}{\fsample} \frac1\pi
        (2\pi \fknee)^{-\gamma} \sin[(\gamma+2)\pi/2]
\eeq
to obtain the desired power spectrum  
\beq
      P(f) = \frac{\sigma^2}{\fsample} (\frac{f}{\fknee})^\gamma
               \qquad(\fmin\ll f\ll\fmax).   \label{apowerlaw}
\eeq
Here $\sigma$ is the white noise std, $\fsample$ is the sampling frequency,
and $\fknee$ is the knee frequency, at which the power of the 
power-law noise component equals the white noise power 
$\sigma^2/\fsample$.
The maximum $\fmax$ 
should well exceed the knee frequency.
 
The coefficients $b_k$ are given by
\beq
   b_k = \frac{\sigma^2}{\fsample} \frac1\pi
        (2\pi\fknee)^{-\gamma} \sin[(\gamma+2)\pi/2]
        g_k^{\gamma+1}\Delta
\eeq
for $-2<\gamma<0$.
Especially, for $1/f$ noise ($\gamma=-1$) we have the simple 
formula
\beq
     b_k = 2\sigma^2\frac{\fknee}{\fsample} \Delta. 
\eeq 

The case $\gamma=-2$ requires a separate treatment. 
We see directly from Eq. (\ref{aoneprocess}), that the 
desired spectrum is given by one single $g$ 
component with coefficient
\beq
     b = \frac{2\pi^2\fknee^2}{g}\frac{\sigma^2}{\fsample}.
\eeq
The spectrum then has the form
\beq
    P(f) = \frac{\fknee^2}{\fmin^2+f^2}
             \frac{\sigma^2}{f_{\rm s}} 
\eeq
where $\fmin=g/(2\pi)$. The spectrum behaves as $\propto f^{-2}$ at 
$f\gg\fmin$ and levels out below $\fmin$.


\subsection{Component covariance matrices}
\label{app:covcomp}

Once the expansion (\ref{aeexpansion}) has been found,
the covariance matrix can be computed as
\beq
   \bC_a = \sum_kb_k\bC^k.
\eeq
In the following we calculate the component covariance matrices 
$\bC^k$.

In Section 4 we derived the general formula
\bea
   \lefteqn{C_{ii'll'}=}  \label{agencov} \\ 
    && \sum_{j,j'=0}^{n-1}\bar F_{lj}\bar F_{l'j'} 
     \frac{1}{M^2}\sum_{m,m'=0}^{M-1}  
      c_{[(i-i')nM+(m-m')n+j-j']}.     \nonumber        
\eea

Assume then that the auto-correlation 
function is of the exponential form
\beq
    \langle y_py_{p'}\rangle
    = c_{k,p-p'} = \exp(-g_k|t|) 
     = \exp\big(-\frac{g_k}{\fsample}|p-p'|\big)
\eeq
where $\fsample$ is the sampling frequency and indices $p,p'$ label 
samples along the TOD. We substitute this into Eq. (\ref{agencov}).
The covariance becomes 
\bea
    C^k_{ii'll'} &=&                   
      \sum_{jj'=0}^{n-1}\bar F_{lj}\bar F_{l'j'}
      \frac{1}{M^2}\sum_{m,m'=0}^{M-1}   \label{amultisum}  \\
    &&\times\exp\big(-\frac{g_k}{\fsample}|(i-i')nM+(m-m')n+j-j'| 
                                     \big).  \nonumber
\eea
We treat cases $|i-i'|>0$ and $i=i'$ separately.

\begin{enumerate}

\item $|i-i'|>0$. 
If $i-i'>0$ the quantity inside the brackets is positive, 
and we can split the four-dimensional sum into a product of four 
sums as
\bea
   C^k_{ii'll'} &=&
   \exp\big(-\frac{g_k}{\fsample}(i-i'-1)nM \big)        \nonumber  \\  
   &&\times\frac{1}{M^2} \sum_{m=0}^{M-1}{\rm e}^{-\frac{g_k}{\fsample}mn} 
                    \sum_{m=0}^{M-1}{\rm e}^{-\frac{g_k}{\fsample}(M-1-m)n} 
                                                   \nonumber  \\
   &&\times \sum_{j=0}^{n-1}\bar F_{lj}{\rm e}^{-\frac{g_k}{\fsample}j}
            \sum_{j'=0}^{n-1}\bar F_{l'j'}
             {\rm e}^{-\frac{g_k}{\fsample}(n-j')}.   \label{acase1}        
\eea
We have arranged the terms in such a way that the argument of an 
exponent function is always negative. This is helpful in 
numerical evaluation.
The sum over $m,m'$ can be carried out analytically, yielding
\bea
   G_k &=& \frac{1}{M^2} \sum_{m=0}^{M-1}{\rm e}^{-\frac{g_k}{\fsample}mn} 
                    \sum_{m=0}^{M-1}{\rm e}^{-\frac{g_k}{\fsample}(M-1-m)n} 
                                                   \nonumber  \\                  
   &=& \frac{1}{M^2} \frac{(1-{\rm e}^{-\frac{g_k}{\fspin}M})^2}
                     {(1-{\rm e}^{-\frac{g_k}{\fspin}})^2}  
\eea
where $\fspin=\fsample/n$.

The elements for which $i-i'<0$ are obtained from the symmetry 
relation,$C^k_{ii'll'}=C^k_{i'il'l}$.

\item The case $i=i'$ is more complicated, 
since the quantity inside the brackets in Eq. (\ref{amultisum}) 
takes both positive and negative values. 
We split the sum into three terms 
($m=m'$, $m>m'$, and $m<m'$) and the $m=m'$ term further 
into three terms 
($j=j'$, $j>j'$, and $j<j'$), 
\bea
     C^k_{iill'} &=&               
      \sum_{jj'=0}^{n-1}\bar F_{lj}\bar F_{l'j'}
      \frac{1}{M^2}\sum_{m,m'=0}^{M-1}\  
               {\rm e}^{-\frac{g_k}{\fsample}|(m-m')n+j-j'| }
            \nonumber    \\
      &=&\frac{1}{M^2}\sum_{m=0}^{M-1}\sum_{m'<m}
             {\rm e}^{-\frac{g_k}{\fsample}(m-m'-1)n}         \nonumber  \\
     &&     \quad\times\sum_{j=0}^{n-1}\bar F_{lj}
              {\rm e}^{-\frac{g_k}{\fsample}j}
             \sum_{j'=0}^{n-1}\bar F_{l'j'}
             {\rm e}^{-\frac{g_k}{\fsample}(n-j')}             \nonumber  \\
     &+& \frac{1}{M^2}\sum_{m'=0}^{M-1}\sum_{m<m'}
             {\rm e}^{-\frac{g_k}{\fsample}(m'-m-1)n}          \nonumber \\
     &&   \quad\times  \sum_{j=0}^{n-1}\bar F_{lj}
              {\rm e}^{-\frac{g_k}{\fsample}(n-j)}
             \sum_{j'=0}^{n-1}\bar F_{l'j'}
             {\rm e}^{-\frac{g_k}{\fsample}j}                 \nonumber   \\
     &+& \frac1M  \sum_{j=0}^{n-1}\bar F_{lj} \bar F_{l'j} 
       + \frac1M  \sum_{j=0}^{n-1}\bar F_{lj} \sum_{j'<j}\bar F_{l'j'}
          {\rm e}^{-\frac{g_k}{\fsample}(j-j')}              \nonumber \\
     &+& \frac1M  \sum_{j'=0}^{n-1}\bar F_{l'j'} \sum_{j<j'}\bar F_{lj} 
           {\rm e}^{-\frac{g_k}{\fsample}(j'-j)}.   \label{acase2}      
 \eea
The sum over $m,m'$ can again be calculated analytically,
\bea
     G_k^0 &=& \frac{1}{M^2}\sum_{m=0}^{M-1}\sum_{m'<m}
             \exp(-\frac{g_k}{\fspin}(m-m'-1))     \label{aGk0} \\
     &=& \frac1M\frac{1}{1-\exp(-\frac{g_k}{\fspin})}
     \Big[ 1-\frac1M\frac{1-\exp(-\frac{g_k}{\fspin}M)}
                         {1-\exp(-\frac{g_k}{\fspin})} 
                                       \Big].   \nonumber
\eea
Formula (\ref{acase2}) may seem complicated, but is easy and fast
to evaluate numerically.

\end{enumerate}

Equations (\ref{acase1}) and (\ref{acase2}), when written in compact form, 
give 
the formulae (\ref{covpos})-(\ref{covzero}) in Section 4.

\clearpage


\begin{thebibliography}{}

\bibitem[\protect\citeauthoryear{Burigana}{1997}]{Burigana97}
Burigana C., Malaspina M., Mandolesi N., Danese L., Maino D., 
Bersanelli M., Maltoni M., 1997,
Int. Rep. TeSRE/CNR, 198/1997, November, (astro-ph/9906360)

\bibitem[\protect\citeauthoryear{Delabrouille}{1998}]{Delabrouille98} 
Delabrouille J., 1998, A\&AS, 127, 555

\bibitem[\protect\citeauthoryear{Maino}{1999}]{Maino99} 
Maino D. et al., 1999, A\&AS, 140, 383

\bibitem[\protect\citeauthoryear{Maino}{2002}]{Maino02} 
Maino D., Burigana C., G\'orski K.M., Mandolesi N., Bersanelli M., 
2002, A\&A, 387, 356

\bibitem[\protect\citeauthoryear{Keih\"anen}{2004}]{Keihanen04} 
Keih\"anen E., Kurki-Suonio H., Poutanen T., Maino D., Burigana C., 
2004, A\&A, 428, 287

\bibitem[\protect\citeauthoryear{Natoli}{2001}]{Natoli01} 
Natoli P., de Gasperis G., Gheller C., Vittorio N., 2001, 
A\&A, 372, 346

\bibitem[\protect\citeauthoryear{Dor\' e}{2001}]{Dore01} 
Dor\'e O., Teyssier R., Bouchet F.R., Vibert D., Prunet S. 2001, 
A\&A, 374, 358

\bibitem[\protect\citeauthoryear{Borrill}{1999}]{Borrill99} 
Borrill, J. 1999, preprint (astro-ph/9911389)

\bibitem[\protect\citeauthoryear{G\'orski}{1999}]{Gorski99} 
G\'orski K.M., Hivon E., Wandelt B.D., 1999,
in Proceedings of the MPA/ESO Cosmology Conference "Evolution of
Large-Scale Structure", ed. A.J. Banday, R.S. Sheth, \& L. Da
Costa, PrintPartners Ipskamp, NL, 37, (astro-ph/9812350)

\bibitem[\protect\citeauthoryear{G\'orski}{2004}]{Gorski04} 
G\'orski K.M., Hivon E., Banday A.J., Wandelt B.D., Hansen F.K., 
Reinecke M., Bartelman M., 2004, preprint (astro-ph/0409513)

\bibitem[\protect\citeauthoryear{Press}{1992}]{NumRes} 
Press W.H., Teukolsky S.A., Vetterling W.T., Flannery B.P., 1992,
Numerical Recipes, 2nd ed. (Cambridge University Press, Cambridge)

\bibitem[\protect\citeauthoryear{Seljak}{1996}]{cmbfast96} 
Seljak U., Zaldarriaga M., 1996, ApJ, 469, 437

\end{thebibliography}
\end{document}